\documentclass[a4paper,11pt]{article}
\pdfoutput=1
\usepackage{jheppub}
\usepackage{amsfonts,amssymb,amsmath}
\usepackage[normalem]{ulem}
\usepackage{color}
\usepackage{epstopdf}
\usepackage{tensor}
\usepackage{enumerate}

\newcommand{\be}{\begin{equation}}
\newcommand{\ee}{\end{equation}}
\newcommand{\ba}{\begin{eqnarray}}
\newcommand{\ea}{\end{eqnarray}}
\newcommand{\beal}{\begin{aligned}}
\newcommand{\eeal}{\end{aligned}}
\newcommand{\nn}{\nonumber}

\title{Generalized quasi-topological gravity}

\author[a]{Robie A. Hennigar}
\author[b,a]{David %``The Living Legend" 
Kubiz\v n\'ak}
\author[a,b]{Robert %BatMann
B. Mann}

\affiliation[a]{Department of Physics and Astronomy, University of Waterloo,
Waterloo, Ontario, Canada, N2L 3G1}
\affiliation[b]{Perimeter Institute, 31 Caroline Street North, Waterloo,
ON, N2L 2Y5, Canada}

\emailAdd{rhennigar@uwaterloo.ca}
\emailAdd{dkubiznak@perimeterinstitute.ca}
\emailAdd{rbmann@uwaterloo.caa}

\abstract{
We construct the most general, to cubic order in curvature, theory of gravity whose (most general) static spherically symmetric vacuum solutions are fully described by a single field equation. The theory possess the following remarkable properties: i) it has a well-defined Einstein gravity limit ii) it admits `Schwarzschild-like' solutions characterized by a single metric function iii) on maximally symmetric backgrounds it propagates the same degrees of freedom as Einstein's gravity iv) Lovelock and quasi-topological gravities, as well as the recently developed Einsteinian cubic 
 gravity~\href{http://arxiv.org/abs/1607.06463}{{\tt
  ArXiv:1607.06463}} in four dimensions, are recovered as special cases.
We perform a brief analysis of asymptotically flat black holes in this theory and study their thermodynamics.
}

\preprint{DCPT-17/03}

\begin{document}

\maketitle

\section{Introduction}

Higher derivative models of gravity are ubiquitous in high energy physics.  It was realized more than forty years ago that the inclusion of quadratic terms in the gravitational action can lead to a power counting renormalizable theory of gravity~\cite{Stelle:1976gc}. Higher order interactions are generically expected to appear in the low energy effective action of whichever theory provides the UV completion to General Relativity.  The types of interactions generated depends on the model considered: the Gauss-Bonnet term, which falls into the Lovelock class~\cite{Lovelock:1971yv}, appears in the low energy limit of the heterotic superstring~\cite{Zwiebach:1985uq}, while various curvature cubed terms appear in the $\alpha'^2$ corrections to the bosonic string~\cite{Metsaev:1986yb}.

Higher curvature gravity has also played a prominent r\^ole in holography.  For example, holography has been used to impose physical bounds on the couplings of various higher curvature theories by imposing consistency of the dual CFT (e.g. causality)~\cite{Camanho:2009vw, Buchel:2009sk, Camanho:2009hu,Hofman:2009ug}.  The inclusion of quadratic terms has been shown to lead to violations of  the Kovtun--Son--Starinets (KSS) viscosity/entropy ratio bound~\cite{Kovtun:2004de, Brigante:2007nu}.  Holography has even motivated the construction of new  higher curvature gravities, such as quasi-topological gravity~\cite{Oliva:2010eb, Oliva:2010zd, Myers:2010ru, Myers:2010jv,Cisterna:2017umf, Ghodsi:2017iee}.

Quasi-topological gravity  is an intriguing theory since, when evaluated on spherically symmetric backgrounds, the field equations (which on generic backgrounds are fourth order) reduce to second order differential equations and admit exact solutions of a form very similar to   Lovelock gravity.  However,  for a given order in the curvature, the quasi-topological term is gravitationally active in  dimensions smaller than is the case for the corresponding Lovelock term. For example, the cubic quasi-topological term acts non-trivially in five and higher dimensions, as opposed to seven and higher dimensions for cubic Lovelock gravity.  A further interesting feature of quasi-topological gravity is that on maximally symmetric backgrounds the linearized equations of motion coincide with the linearized Einstein equations up to an overall prefactor.  The advantages of this are twofold.  First, it avoids a feature that makes many higher curvature theories physically unpalatable: propagating, on the vacuum, extra degrees of freedom beyond  those present in Einstein gravity, some of which carry negative kinetic energy---\textit{ghosts}---which is equivalent to a breakdown of unitarity in the quantum theory~\cite{Sisman:2011gz}.  Second, since the structure of the graviton propagator is the same as that in Einstein gravity, holographic studies of the theory are significantly simplified~\cite{Myers:2010jv, Myers:2010tj}.

Recently, a new cubic curvature interaction---named \textit{Einsteinian cubic gravity}---has been found which, like the Lovelock theory, matches the Einstein equations at the linear level and has the coefficients appearing in the action `independent of dimension'~\cite{Bueno:2016xff}. 
  Contrary to the cubic Lovelock or quasi-topological terms, this new term is non-trivial in four-dimensions.  Studies of black holes of the theory~\cite{Hennigar:2016gkm, Bueno:2016lrh} revealed that, in four dimensions, the theory admits {\em static spherically symmetric} (SSS) solutions, {in {\em vacuum} (VSSS) or in the presence of suitable matter (e.g. a Maxwell field)}, of the form
\be\label{eqn:metricAnsatz}
ds^2 = - N^2 f dt^2 + \frac{dr^2}{f} + r^2 d \Sigma^2_{(d-2), k}\,,
\ee
with $N = const.$, i.e., the solution  is characterized in terms of a single metric function $f$. (Henceforth  we normalize $N$ to unity, setting $N=1$. This can be achieved without loss of generality by reparametrizing the time coordinate $t$.) Here $d \Sigma^2_{(d-2), k}$ is the line element on a surface of constant scalar curvature $k = +1,0, -1$ corresponding to spherical, flat, and hyperbolic topologies.  As remarked in \cite{Bueno:2016lrh},  Einsteinian cubic gravity in $d=4$ dimensions is \textit{the most general theory} (up to cubic order) which admits a {\em Schwarzschild-like} solution, that is, a SSS solution characterized by a single metric function in vacuum or in the presence of suitable matter.
However, this is no longer true in higher dimensions  where the choice $N=1$ leads to inconsistent field equations \cite{Hennigar:2016gkm}.

 It is then natural to ask, what is the most general theory of gravity in $d>4$ dimensions for which the most general VSSS takes the form \eqref{eqn:metricAnsatz} with $N=1$. This is certainly true for the Lovelock and quasi-topological gravities to all orders and in any number of dimensions.  However, as we shall see, this is not the end of the story.

 The aim of this paper is to answer the above question to cubic order in curvature.
In fact, we impose a slightly stronger condition, and ask the following: ``{\em What is the most general cubic theory for which the VSSS is fully characterized by a single field equation?}'' This condition requires that the time and radial components of the field equations for the ansatz \eqref{eqn:metricAnsatz} are equal,
\be\label{EttErr}
\mathcal{E}_{t}^{t}  = \mathcal{E}_{r}^{r}\,.
\ee
The equality is required up to terms that vanish upon setting $N=1$ and should be valid off-shell, that is, for any function $f$ in \eqref{eqn:metricAnsatz}.

We will show that the most general $d$-dimensional theory (up to and including cubic order in curvature) obeying \eqref{EttErr} is given by
\be\label{eqn:actionWithS}
\mathcal{I} = \frac{1}{16 \pi G} \int d^d x \sqrt{-g} \bigl[-2 \Lambda + R + \alpha \mathcal{X}_4+  \beta \mathcal{X}_6 + \mu \mathcal{Z}_d -  \lambda \mathcal{S}_d \bigr]\,.
\ee
Here, $\Lambda$ is the cosmological constant and $\alpha,\beta,\mu, \lambda$ are arbitrary coupling constants. $R$ stands for the Ricci scalar and $\mathcal{X}_4$ and $\mathcal{X}_6$ are the four- and six-dimensional Euler densities,
\ba\label{X6}
\mathcal{X}_4&=&-\frac{1}{4}\delta^{a_1b_1a_2b_2}_{c_1d_1c_2d_2}R_{a_1b_1}{}^{c_1d_1}R_{a_2b_2}{}^{c_2d_2}\,,\nonumber\\
\mathcal{X}_6  &=& - \frac{1}{8}
\delta^{a_1b_1a_2b_2a_3b_3}_{c_1d_1c_2d_2c_3d_3}R_{a_1b_1}{}^{c_1d_1}R_{a_2b_2}{}^{c_2d_2}R_{a_3b_3}{}^{c_3d_3}\,,
\ea
where $\delta_{c_{1}d_{1}\ldots c_{k}d_{k}}^{a_{1}b_{1}\ldots a_{k}b_{k}}$ are the generalized Kronecker delta functions, totally antisymmetric in both sets of indices, and $R_{a_{k}b_{k}}^{\quad c_{k}d_{k}}$  is the Riemann tensor\footnote{Note that we are using the opposite sign of the normalization used in~\cite{Myers:2010ru} for the Euler densities.}.
$\mathcal{Z}_d$ is the cubic quasi-topological term given by \eqref{Zd} below,  and $\mathcal{S}_d$ is a hitherto unnoticed term whose explicit form reads
\ba\label{Sd}
\mathcal{S}_d &=&
14 R_{a}{}^{e}{}_{c}{}^{f} R^{abcd} R_{bedf}+ 2 R^{ab} R_{a}{}^{cde} R_{bcde}- \frac{4 (66 - 35 d + 2 d^2) }{3 (d-2) (2 d-1)} R_{a}{}^{c} R^{ab} R_{bc}\nonumber\\
&& -  \frac{2 (-30 + 9 d + 4 d^2) }{(d-2) (2 d-1)} R^{ab} R^{cd} R_{acbd} -  \frac{(38 - 29 d + 4 d^2)}{4 (d -2) (2 d  - 1)} R R_{abcd} R^{abcd}  \nonumber\\
&&+ \frac{(34 - 21 d + 4 d^2) }{(d-2) ( 2 d - 1)} R_{ab} R^{ab} R -  \frac{(30 - 13 d + 4 d^2)}{12 (d-2) (2 d - 1)}  R^3\,.
\ea

The new theory \eqref{eqn:actionWithS} possesses a number of remarkable properties.
\begin{enumerate}[i)]

\item  Upon setting the couplings $\alpha,
\beta, \mu$ and $\lambda$ to zero, one recovers Einstein gravity with a cosmological constant.

\item In any number of dimensions, the theory \eqref{eqn:actionWithS} admits the  `Schwarzschild-like' VSSS solutions
 (given by \eqref{eqn:metricAnsatz} with $N=1$) characterized by a single metric function.  This is also true in the presence of suitable matter. In particular, in four dimensions the new term
$\mathcal{S}_d$ reduces to Einsteinian cubic gravity term plus an additional term which vanishes on a VSSS ansatz \eqref{eqn:metricAnsatz}.  This confirms the claim made in~\cite{Bueno:2016lrh} that the Einsteinian cubic gravity is the most general four-dimensional theory with this property.

\item  On a maximally symmetric background, the action \eqref{eqn:actionWithS}
leads to a theory whose linearized equations coincide with the linearized Einstein equations, which means that (on these backgrounds) the theory
propagates the same degrees of freedom as Einstein's gravity.

\item  The single non-linear vacuum field equation for the SSS \eqref{eqn:metricAnsatz} is of third-order for $f$ and takes a form of a total derivative, effectively reducing to a second order differential equation upon a trivial integration in vacuum or in the presence of suitable matter. For vanishing $\lambda$ we obtain second-order differential equations only.
\end{enumerate}

In some sense the preceding statement is equivalent to the statement that we seek the most general cubic curvature theory with a well defined Einstein limit admitting a VSSS solution with $N=1$. However before moving on to the construction we add some qualification to the statement \textit{most general theory}.  We are considering additional higher curvature terms supplementing the Einstein-Hilbert action in a manner such that these terms can be ``turned off" by a suitable adjustment of parameters in the action.  Our conditions are the same as those mentioned in~\cite{Bueno:2016lrh}:
\begin{enumerate}
\item The solution is not an `embedding' of an Einstein gravity black hole into a higher order gravity~\cite{delaCruzDombriz:2009et, Smolic:2013gz, Lu:2012xu}.  That is, the solution must be modified by the addition of the higher curvature terms.

\item The solution is not of a pure higher order gravity, but includes the Einstein-Hilbert term.  For example, pure Weyl-squared gravity allows for four dimensional solutions with $N=1$~\cite{delaCruzDombriz:2009et, Riegert:1984zz, Oliva:2011xu, Oliva:2012zs}.

\item Further, the theory must admit an Einstein-Gravity limit, i.e. reduce to the Einstein-Hilbert action upon setting
some of the parameters in the action to zero.  This excludes certain theories that tune the couplings between the various orders of curvature terms \cite{Bueno:2016dol, Cai:2009ac}.

\end{enumerate}

The remainder of the paper is organized as follows.  We first present the construction of the theory and study its linearization on a maximally symmetric background.  We then move on to presenting the non-linear field equations and study their asymptotically flat black hole solutions.

%%%%%%%%%%%%%%%%%%%%%%%%%%%%%%%%%%%%%%%%%%%%
%%%%%%%%%%%%%%%%%%%%%%%%%%%%%%%%%%%%%%%%%%%%%%
\section{Construction of the theory}

We seek a higher curvature theory which for a SSS ansatz \eqref{eqn:metricAnsatz} yields a single field equation, that is \eqref{EttErr} is required, up terms that vanish upon setting $N=1$, for any function $f$.
Up to quadratic order, the answer is simple: {\em the most general theory of this nature is Einstein gravity supplemented by the Gauss-Bonnet term.}

Considering next cubic interactions, there are twelve non-vanishing cubic densities. However, relationships between these reduce the total number to ten when a total derivative is discarded~\cite{Myers:2010ru}. Thus we employ the following basis of the ten cubic densities:
\ba\label{terms}
\mathcal{L}_1 &=& R_{a}{}{^c}_{b}{}^{d}R_c{}^e{}_d{}^f R_e{}^a{}_f{}^b \, ,
\quad
\mathcal{L}_2 = R_{ab}{}^{cd} R_{cd}{}^{ef} R_{ef}{}^{ab} \, ,\quad
\mathcal{L}_3 = R_{abcd}R^{abc}{}_e R^{de} \,,\nn\\
\mathcal{L}_4 &=& R_{abcd}R^{abcd}R \,,\quad
\mathcal{L}_5 =  R_{abcd}R^{ac}R^{bd}\, ,
\quad
\mathcal{L}_6 = R_a{}^b R_b{}^c R_c{}^a \, ,
\nn\\
\mathcal{L}_7 &=& R_a{}^b R_b{}^a R \, ,
\quad
\mathcal{L}_8 = R^3 \, , \quad \mathcal{L}_{9} = \nabla_a R_{bc} \nabla^a R^{bc}\,,\quad
\mathcal{L}_{10} = \nabla_a R \nabla^a R \, .
\ea
With these terms, we write the action for the theory as
\be
\mathcal{I} = \frac{1}{16 \pi G} \int d^d x \sqrt{-g} \Bigl(R - 2 \Lambda +   \sum_i c_i  \mathcal{L}_i \Bigr)\,,
\ee
where $c_i$ are constants to be constrained by the condition \eqref{EttErr}, for the
generalized Einstein tensor
\be
\mathcal{E}_{ab} = \frac{1}{\sqrt{-g}} \frac{\delta \mathcal{I}}{\delta g^{ab}}\,.
\ee

Let us now sketch out the procedure schematically.  There are two methods by which one may arrive at the theory of interest.  The first involves computing the field equations of the theory.  The relationship between the $tt$ and $rr$ field equations can be written in the following (schematic) form:
\be
\mathcal{E}_{t}^{t} = \mathcal{E}_{r}^{r}+\mathcal{E}_0(f,f',\dots)  + N\mathcal{E}_1(f,f',\dots)+N'\mathcal{E}_2(f,f',\dots)+N''\mathcal{E}_3(f,f',\dots)+\cdots  
\ee
where the `$+ \cdots$'  indicate terms which may contain higher derivatives or different powers of $N$.
Condition $\eqref{EttErr}$ with $N = 1$ will be satisfied provided
 the terms that are not multiplied by a derivative of $N$, $\mathcal{E}_0$ and $\mathcal{E}_1$, vanish for any $f$.  This leads to a set of algebraic equations for the $c_i$'s which can be solved to yield the relevant theory.

The second method makes use of the Weyl method~\cite{palais1979, Deser:2003up}: substituting the metric ansatz~\eqref{eqn:metricAnsatz} directly into the action and varying with respect to $N$ and $f$ to obtain the field equations.  From this perspective, the theory will admit a solution with $N =1$ provided that upon  repeated integration by parts, the action can be put into the form
\be\label{eqn:requiredActionForm}
\mathcal{I} = \frac{1}{16 \pi G} \int dr N(r) F' + \left\{ \, \cdots \right \}
\ee
where we have suppressed the angular part and the dots represent terms containing two or more powers of derivatives  of $N$---e.g. $N'^2/N$, $N'N''/N$ and so on---such terms cannot be eliminated by further integration by parts.  The crucial feature is that the term multiplying $N(r)$ is a total derivative  and is a functional only of $f$ and its derivatives. This guarantees that performing a variation of this term with respect to $f$ yields an expression containing only derivatives of $N$, which can then be solved by setting $N = 1$.  By demanding the form of the action shown above, one can construct algebraic equations which constrain the constants $c_i$.

In what follows we will employ the first method and then cross-check it using the second method.  Let us first discuss the situation in $d > 4$ dimensions and then turn to the (slightly different) four-dimensional case.

%%%%%%%%%%%%%%%%%%%%%%%%%%%%%%%%%%%%%%
\subsection*{$d>4$ dimensions}

We find that only three couplings $c_i$ are independent; we choose these to be $c_1, c_2$ and $c_3$, while the others are constrained in the following way\footnote{In practice, we evaluated the field equations explicitly using \textsc{Mathematica} for $d=5$ to $d=9$ and inferred the dimension dependence from these results.  We subsequently verified and cross-checked the result using \textsc{Maple} up to $d=19$ (see later discussion).}:
\ba\label{ccs2}
c_4 &=& \frac{3d-6}{8(d-2)(2d-1)} c_1 + \frac{6 + 6d - 3d^2}{2(d-2)(2d-1)} c_2 + \frac{1 + 2d - d^2}{2(d-2)(2d-1)} c_3\,,
\nn\\
c_5 &=& {\frac {6-3\,d}{ \left( d-2 \right)  \left( 2\,d-1 \right) }} c_1 + {\frac {-48+36\,d-12\,{d}^{2}}{ \left( d-2 \right)  \left( 2\,d-1
 \right) }} c_2 + -\,{\frac {4(3-3\,d+{d}^{2})}{ \left( d-2 \right)  \left( 2\,d-1
 \right) }} c_3\,,
\nn\\
c_6 &=& \frac{4}{2d-1} c_1 +  \,{\frac {8(8-5\,d)}{ \left( d-2 \right)  \left( 2\,d-1 \right) }} c_2 + \,{\frac {2(18-7\,d-2\,{d}^{2})}{3 \left( d-2 \right)  \left( 2\,d-1
 \right) }} c_3\,,
 \nn\\
c_7 &=&  \,{\frac {6-3\,d}{2 \left( d-2 \right)  \left( 2\,d-1 \right) }} c_1 + {\frac {2(-12+6\,d+3\,{d}^{2})}{ \left( d-2 \right)  \left( 2\,d-1
 \right) }} c_2 + {\frac {2({d}^{2}-2)}{ \left( d-2 \right)  \left( 2\,d-1 \right) }} c_3\,,
\nn\\
c_8 &=&  \frac{1}{8(2d-1)} c_1 + {\frac {6-6\,d-{d}^{2}}{2 \left( d-2 \right)  \left( 2\,d-1
 \right) }} c_2 -{\frac { \left( d-1 \right)  \left( d+3 \right) }{ 6\left( d-2
 \right)  \left( 2\,d-1 \right) }} c_3\,,\nn\\
 c_9&=&0\,,\quad c_{10}=0\,.
\ea
Since there are three free parameters, $\{c_1, c_2, c_3\}$, the resulting theory is a linear combination of three independent cubic densities.  Since both the six-dimensional Euler density and the quasi-topological terms possess this property, there must be choices for the constants which produce these densities.  We find that choosing
\be
c_1 =8 \,, \quad c_2 = -4\,, \quad c_3 = 24\,,
\ee
produces the six-dimensional Euler density, $\mathcal{X}_6$, \eqref{X6}, while
choosing
\be\label{Zd}
c_1 = 1 \, , \quad c_2 = 0 \,, \quad c_3 = - \frac{3(d-2)}{(2d-3)(d-4)}\,,
\ee
produces the quasi-topological density
\ba
\mathcal{Z}_d &=&    {{{R_a}^b}_c{}^d} {{{R_b}^e}_d}^f {{{R_e}^a}_f}^c
               + \frac{1}{(2d - 3)(d - 4)} \Bigl( \frac{3(3d - 8)}{8} R_{a b c d} R^{a b c d} R  - \frac{3(3d-4)}{2} {R_a}^c {R_c}^a R \nonumber \\
              &&  - 3(d-2) R_{a c b d} {R^{a c b}}_e R^{d e} + 3d  R_{a c b d} R^{a b} R^{c d}
                + 6(d-2) {R_a}^c {R_c}^b {R_b}^a  + \frac{3d}{8} R^3 \Bigr)\,.\quad
\label{Zd}
\ea
Note that the expression $\mathcal{Z}'_d$ from~\cite{Myers:2010ru} can be obtained by choosing
\be
c_1 = 0 \, , \quad c_2 = 1 \,, \quad c_3 = - \frac{12(d^2-5d + 5)}{(2d-3)(d-4)} \, .
\ee
 However, this term is not independent from the quasi-topological term and the six-dimensional Euler density, but rather~\cite{Myers:2010ru}
\be
\mathcal{X}_6 = -4\mathcal{Z}'_d + 8\mathcal{Z}_d\,,
\ee
and so $\mathcal{Z}'_d$ is not the third invariant we are looking for.

Since there is an extra free parameter, we obtain an additional independent cubic density which shares the property of permitting a solution with a single metric function.  We find that this term cannot be expressed as a linear combination of the five invariants introduced in~\cite{Oliva:2010zd}; though this is not surprising since here the traced field equations are of the fourth order rather than third.  The new term could be obtained by setting $c_3=1$ and $c_1=c_2=0$, but as we will see shortly a more convenient choice (motivated by the Einsteinian theory in four dimensions) is
\be\label{eqn:newTermCoeffs}
c_1 = 14 \,, \quad c_2 = 0 \,, \quad c_3= 2\,,
\ee
which recovers $S_d$ given by \eqref{Sd}.

%%%%%%%%%%%%%%%%%%%%%%%%%%%%%%%%%
\subsection*{$d=4$ dimensions}

We now turn to the case of four dimensions.  In four dimensions, demanding $\mathcal{E}_{t}^{t}=\mathcal{E}_{r}^{r}$ leads to 4 independent couplings
and hence to four possible invariants. Choosing $\{c_1, c_2, c_3, c_6\}$, the others are given by the following constraints:
\ba
c_4 &=& \frac{3c_1 - 36 c_2 - 14 c_3}{56}\,,\quad
c_5 = -\frac{3c_1 + 48 c_2 + 14 c_3}{7}\,,\nn\\
c_7 &=& \frac{6c_1 + 96 c_2 + 14 c_3 - 21 c_6}{28}\,,\quad
c_8 = \frac{-3c_1 - 20 c_2 +  7 c_6}{56}\,,\nonumber\\
c_9&=&0\,,\quad c_{10}=0\,.
\ea
We find that the choice
\be
c_1 = 8 \, , \quad c_2 = - 4 \, , \quad c_3 = 24 \, , \quad c_6 = -16 \, .
\ee
gives the six-dimensional Euler density, which vanishes identically in $d < 6$ due to the Schouten identities.  The remaining three interactions are given by the following convenient choices of coefficients:
\begin{align}
\mathcal{P} &: \quad  c_1 = 12 \,, \quad \! \! \! c_2 = 1 \,, \quad c_3 =0 \,, \quad \!\!\! c_6 = 8 \, .
\nn\\
\mathcal{C} &: \quad c_1 =0 \,, \quad c_2 = 0 \, , \quad c_3 = 1 \,, \quad c_6 = 0 \, .
\nn\\
\mathcal{C}' &: \quad c_1 =0 \,, \quad c_2 = 0 \, , \quad c_3 = 0 \,, \quad c_6 = 1 \, .
\end{align}
Here the various terms are given by the following cubic densities:
\ba\label{eqn:topoterms}
\mathcal{P} &=& 12  R_a{}^c{}_b{}^e R_c{}^m{}_e{}^n R_m{}^a{}_n{}^b + R_{ab}{}^{ce} R_{ce}{}^{mn}R_{mn}{}^{ab}
- 12 R_{abcd}R^{a c}R^{be}
+ 8 R_a^b R_b^c R_c^a \,,
\nn\\
\mathcal{C} &=& \tfrac{1}{2} R_{a}{}^{b} R_{b}{}^{a} R - 2 R^{ac} R^{bd} R_{abcd} -  \tfrac{1}{4} R R_{abcd} R^{abcd}
+ R^{de} R_{abcd} R^{abc}{}_{e} \, ,
\nn\\
\mathcal{C}' &=& R_{a}{}^{b} R_{b}{}^{c} R_{c}{}^{a} -  \tfrac{3}{4} R_{a}{}^{b} R_{b}{}^{a} R + \tfrac{1}{8} R^3 \,.
\ea
In the above, $\mathcal{P}$ is the recently discovered Einsteinian cubic term~\cite{Bueno:2016xff}, while $\mathcal{C}$ and $\mathcal{C}'$ are two new terms, which when evaluated on a four-dimensional VSSS metric do not produce any non-trivial contributions to the field equations.  Thus it follows that Einsteinian cubic gravity is the most general cubic theory which admits a single metric function spherical solution.

Now the choice of coefficients used in defining $\mathcal{S}_d$ can be explained.  The choices~\eqref{eqn:newTermCoeffs} lead to the following relationship in four dimensions,
\be
\mathcal{S}_4 - \frac{1}{4} \mathcal{X}_6 +4\mathcal{C}  =  \mathcal{P} \, .
\ee
Since $\mathcal{X}_6$ vanishes identically in four dimensions and $\mathcal{C}$ makes no contribution to the field equations, we see that in four dimensions the theory given by the Lagrangian density $\mathcal{S}_d$ \eqref{Sd} will yield the field equations that coincide with those for the Eisteinian cubic gravity.  {Note that there is no choice of $c_1$, $c_2$ and $c_3$ such that the theory reduces precisely to Einsteinian cubic gravity in four dimensions.}

%%%%%%%%%%%%%%%%%%%%%%%%%%%%%%%%%%%%%%%%%%%%%%%%%%
%%%%%%%%%%%%%%%%%%%%%%%%%%%%%%%%%%%%%%%%%%%%%%%%%
\section{Linearized field equations}

Let us now turn to a discussion of the linearized theory.
To study the implications of the new term $S_d$, we temporarily switch off both Euler densities as well as the quasi-topological term and consider the following action:
\be\label{Sdtheory}
\mathcal{I} = \frac{1}{16 \pi G} \int d^d x \sqrt{-g} \Bigl(R -2\Lambda -  \lambda \mathcal{S}_d \Bigr)\,,
\ee
where $\lambda \ge 0$ is assumed in the following, and we parameterize the cosmological constant as
\be
\Lambda=-\frac{(d-1)(d-2)}{2L^2}\,.
\ee

To linearize the theory, let us first seek the appropriate `vacuum background'.
It is easy to verify that the Minkowski background provides a vacuum for the theory with $\Lambda=0$. Slightly more generally,
considering a nontrivial $\Lambda$ and imposing the maximal  symmetry condition
\be\label{eqn:background}
R_{abcd}^{[0]} = - \frac{2 \beta}{L^2} \, g^{[0]}_{a[c} g^{[0]}_{d]b}
\ee
upon evaluating the field equations for this choice of background, we find the following constraint:
\be
0 = 1 -  \beta
-  \frac{\lambda}{L^4} \frac{(d-3)(d-6)(184 - 514d + 291 d^2 - 49 d^3 + 4d^4) }{12(d-2)(2d-1)}  \beta^3\,,
\ee
 which determines $\beta$. This  in turn  is related to the effective cosmological constant of the theory
\be
\Lambda_{\rm eff} = -\frac{(d-1)(d-2)\beta}{2L^2} \, .
\ee

Due to the cubic character of this condition, the theory will generically have three distinct vacua, with one having a smooth limit to the Einstein case as $\lambda \to 0$.  Note also that for
\be
\frac{\lambda}{L^4} = \frac{-16(d-2)(2d-1)}{9(d-3)(d-6)(184 - 514d + 291 d^2 - 49 d^3 + 4d^4)}\,,
\ee
the equation reduces to
\be
0=1 - \beta + \frac{4}{27} \beta^3\,,
\ee
which corresponds to a case with a single de Sitter vacuum and two AdS vacua that are degenerate.

 Next we consider a  perturbation around the obtained maximally symmetric backgrounds
\be
g_{ab} = g_{ab}^{[0]} + h_{ab}\,.
\ee
The linear equations of motion can be efficiently obtained using the method introduced in~\cite{Bueno:2016xff, Bueno:2016ypa}---they are given by
\begin{align}
\mathcal{E}_{ab}^L  =  \left[1 + \frac{\lambda}{L^4} \frac{(d-3)(d-6)(184 - 514d + 291 d^2 - 49d^3 + 4 d^4)}{4(d-2)(2d-1)} \beta^2 \right] G_{ab}^L\,,
\end{align}
where $G_{ab}^L$ is the Einstein tensor linearized on the background~\eqref{eqn:background}:
\ba
G^L_{ab} &=& -\frac{1}{2} \Bigl[ \nabla_e \nabla^e h_{ab}   + \nabla_{b}\nabla_{a}h^{c}{}_{c} -  2 \nabla_{c}\nabla_{(a}h_{b)}{}^{c}
  + g^{[0]}_{ab} \left(  \nabla_{d}\nabla_{c}h^{cd} -   \nabla_e \nabla^e h^{c}{}_{c} \right)\nonumber\\
  &&\quad + \frac{(d-1)\beta}{L^2} g_{ab}^{[0]}h^{c}{}_{c}  - \frac{2(d-1) \beta}{L^2} h_{ab}\Bigr]\,.
\ea
 We have cross-checked the results using Mathematica in various dimensions.

Note that the cubic contribution to the linearized field equations vanishes in six dimensions, a feature that holds for any cubic density in six dimensions~\cite{Bueno:2016xff}.  Had we retained both Euler densities and the quasi-topological term, the linearized equations of motion would still match the Einstein equations, but the overall constant would differ.

The remarkable fact that the  linearized equations of motion coincide with the  linearized Einstein equations indicates that the massive and scalar modes are suppressed (i.e. they are infinitely heavy) and the theory propagates the same transverse massless graviton as Einstein's gravity.
This result suggests that it is a necessary condition that the linearized field equations coincide with linearized Einstein gravity in order for a single metric function to be possible under spherical symmetry, and we may elevate this to the level of the following \\
{\bf Conjecture}: \textit{If a geometric theory of gravity satisfies Eq.~\eqref{EttErr} for a VSSS ansatz then the linearized field equations of this theory will coincide with the linearized Einstein equations up to an overall constant multiple.}

 Note that this condition is not sufficient, as there are theories that are equivalent to Einstein's gravity at the linear level but \textit{do not} admit single metric function VSSS:   for example, Einsteinian cubic gravity in dimensions other than four~\cite{Hennigar:2016gkm, Bueno:2016lrh}.  One way to see that this is not a sufficient condition is to note that imposing the  absence of the massive graviton and scalar modes leads to two constraint equations on the $c_i$, however we found that (in $d \ge 5$) there are in fact five constraint equations imposed by demanding $\mathcal{E}_{t}^{t} = \mathcal{E}_{r}^{r}$.  The physical origin of the remaining three constraint equations is not clear.  It would be an interesting line of study to determine if this result extends to higher order gravities, or, if not, determining why the cubic case is special.

To summarize this section, our results indicate that on a maximally symmetric background  the graviton propagator for our theory will be given by the standard Einstein gravity result, up to an overall constant.  This fact should significantly simplify holographic studies of the theory.

%%%%%%%%%%%%%%%%%%%%%%%%%%%%%%%%%%%%%%%%%%%%%%%
%%%%%%%%%%%%%%%%%%%%%%%%%%%%%%%%%%%%%%%%%%%%%%%%%
\section{ Black hole solutions }

We complete this work with a brief study of the vacuum field equations and their  asymptotically flat black hole solutions.  A more detailed account will be presented elsewhere~\cite{HennigarPrep}.  Here we will concern ourselves with only the spherically symmetric solutions; the full field equations, valid for any metric, are presented in the appendix.

 To obtain the explicit field equations under spherical symmetry it is simplest to employ  the Weyl method described above: 
 substituting the metric ansatz~\eqref{eqn:metricAnsatz} directly into the action and varying with respect to $N$ and $f$ to obtain the field equations.  From this perspective, the theory will admit a solution with $N = 1$ provided that by repeatedly integrating by parts the action can be put into the form \eqref{eqn:requiredActionForm}, which ensures that
  performing a variation  with respect to $f$ yields an expression containing only derivatives of $N$, which can then be solved by setting $N = 1$.  A remarkable feature of any such theory is that, since the term multiplying $N$ is a total derivative, the field equations (in vacuum or in the presence of suitable matter)
  will be of reduced order since they can be easily integrated once.

For the VSSS spacetime with $N=1$ the single field equation reads $F'=0$\,, where
\ba\label{eqn:fullEFE}
F&=&(d-2)r^{d-3}\Bigl(k- f + \frac{r^2}{L^2} \Bigr) - \lambda k \frac{2686}{13}  \delta_{d,7}  - \lambda \frac{d-3}{2d-1} \Bigl[ (d^2+5d-15)\Bigl(
 \frac{4}{3}  r^{d-4} f'^3\nn\\
&&- 8 r^{d-5} f f'' \bigl(\frac{rf'}{2} + k - f \bigr)  - 2 r^{d-5} ((d-4)f -2k) f'^2 
 + 8(d-5) r^{d-6} ff'( f - k) \Bigr)
 \nn\\
 &&
 -\frac{1}{3} (d-4) r^{d-7}(k-f)^2  \Bigl( \bigl(-d^4 + \frac{57}{4} d^3 - \frac{261}{4} d^2 + 312 d - 489  \bigr)f \nn\\
 &&+ k\bigl( 129 - 192 d + \frac{357}{4} d^2 - \frac{57}{4} d^3 + d^4 \bigr) \Bigr)  \Bigr]\,.
\ea
By integrating this field equation we obtain
\be\label{EOM}
F=(d-2)C\,,
\ee
where $C$ is an integration constant and the factor of $(d-2)$ has been introduced for later convenience.

 {Let us now assume that there exists a black hole solution to \eqref{EOM}, with a horizon located at $r_+$ given by the `convenient' root of $f(r_+)=0$.
Computing the Iyer-Wald entropy~\cite{Wald:1993nt, Iyer:1994ys}:} 
\be
S = -2 \pi \oint d^{d-2} x \sqrt{\gamma_h} E^{abcd} \hat{\varepsilon}_{ab} \hat{\varepsilon}_{cd} \,,\quad
E^{abcd} = \frac{\partial \mathcal{L}}{\partial R_{abcd}}\,,
\ee
where ${\cal L}$ is the Lagrangian density  (see the appendix for the explicit form of $E^{abcd}$)
and $\hat{\varepsilon}_{ab}$ is the binormal to the horizon, we find
\ba
S &=& \frac{A}{4G}\Bigl[1 - \frac{4\lambda}{r_+^4} \frac{d-3}{2d-1} \Bigl( 8 \pi r_+ (d^2 + 5d - 15)(kT + \pi r_+ T^2)
\nn\\
&&\qquad - \frac{(d-4)(4d^3 -33d^2 + 127 d - 166)k^2}{16}
\Bigr)  \Bigr]\,,
\ea
where $A$ is the area of the event horizon in $d$-dimensions and $T = f'(r_+)/(4\pi)$ is the Hawking temperature.  This result reduces to the expression for the Einsteinian cubic gravity in four dimensions~\cite{Hennigar:2016gkm, Bueno:2016ypa}.  It is a distinguishing feature of the theory including $\mathcal{S}_d$ that the Iyer-Wald entropy includes terms proportional to $f'(r_+)$: these terms are not present in Lovelock and quasi-topological theories.  Here we have substituted the temperature for these factors, since expressing $f'(r_+)$ in terms of $r_+$ involves solving a complicated cubic polynomial (see Eq.~\eqref{eqn:nearHorizon2} below).

We have attempted to solve the field equations exactly in various dimensions without success---except in the trivial case $C=0$ which then corresponds to a maximally symmetric spacetime.  Therefore to study the black hole solutions we work perturbatively.  For convenience, in the present work we shall restrict to the case of {\em asymptotic flatness}, though extensions to other maximally symmetric asymptotics is in progress~\cite{HennigarPrep}.  In what follows we shall keep factors of $k$ visible since they serve as useful accounting devices and  make the generalization to AdS asymptotics easier, setting $k=1$ at the end of the calculation.

 Let us first consider the asymptotic behaviour, $r\to \infty$, treating the $\lambda$ terms as a small correction. We expand the metric function as
\be
f(r) = k - \frac{C}{r^{d-3}} +  \epsilon \, h(r)\,,
\ee
where $\epsilon$ is a parameter used to control the order of $h(r)$, whose contribution is assumed to be small.  We  substitute this into Eq.~\eqref{EOM} and keep terms to order $\epsilon$, i.e. linear in $h(r)$, and then set $\epsilon=1$.  This leads to a second order inhomogeneous differential equation for $h(r)$ which we do not write explicitly here for brevity.
A particular solution, to first order in $\lambda$, takes the form,
\ba
h_p(r) &=& - \frac{2686 k^3 \lambda}{65 r^4} \delta_{d,7} + \frac{ (d-3) \lambda C^2 }{(2d-1)}  \Bigl[ - \frac{2 (d-1)^2 (d^2 + 5d - 15) k}{r^{2d-2}} \nn\\
&&+ \frac{(4d^4 - 3d^3 + 141d^2 - 506 d + 224) C}{4 r^{3d-5}}\Bigr]+
\mathcal{O} \left( \frac{\lambda^2 C}{r^{2d-2}} \delta_{d,7} +    \frac{ \lambda^2 C^3}{r^{3d-1}} \right)\,.
\ea
The presence of the Kronecker delta term for seven dimensions is a curious feature reminiscent of   Chern--Simons gravity~\cite{Crisostomo:2000bb}.  This term will appear in the mass for the seven dimensional black holes, ensuring that the mass vanishes when the horizon radius vanishes, as we shall see in a moment.
 The homogeneous equation reads
\be
h_h'' - \frac{2}{r} h_h' -\omega^2 r^{d-1}  h_h = 0\,,
\ee
where
\be
\omega^2 =  \frac{(2d-1)(d-2)}{4k(d-1)(d-3)(d^2+5d-15) C \lambda}\,.
\ee
{While this equation admits an exact solution in terms of Bessel functions, the relevant details can be captured through an approximate solution for large $r$. In this limit, the} dominant contribution will come from the third term above, and so the first derivative term can be neglected.  Under this approximation,  the solution takes the approximate form\footnote{Note that a more careful analysis is required in the cases where $\omega^2 < 0$.
}
\be
h_h(r) \approx A \exp \left(\frac{\omega r^{(d+1)/2}}{d+1}  \right) + B \exp \left(-\frac{\omega r^{(d+1)/2}}{d+1}  \right) \, .
\ee
Asymptotic flatness demands that $A = 0$, so we are left with the leading order correction,
\be\label{eqn:asympExpansion}
h(r) \approx h_p(r) + B \exp \left(-\frac{\omega r^{(d+1)/2}}{d+1}  \right)  \, .
\ee
Since the homogeneous solution is exponentially suppressed, we are justified in dropping it and considering only the particular solution as the correction.  The presence of the decaying exponential is reminiscent of the Yukawa-type terms which appear in theories containing a massive graviton~\cite{Stelle:1977ry}.  From the linear analysis in the previous section, we know that this theory does not propagate such massive modes on the vacuum---and, indeed, the term vanishes at large $r$.  However, nothing prevents the appearance of these exponential terms on different backgrounds, but note that the falloff is in fact much stronger than the standard Yukawa potential.

We can now compute the ADM mass, which in the asymptotically flat case (with $k=1$) is simply given by~\cite{Deser:2002jk}
\be
\label{eqn:adm_mass}
M = \frac{d-2}{16 \pi G} \omega_{(k)d-2} \lim_{r \to \infty} r^{d-3}(k- g_{tt})
= \frac{d-2}{ 16 \pi G} \omega_{(k)d-2} \left[ C + \frac{2686 k \lambda}{65} \delta_{d,7} \right] \, ,
\ee
where $\omega_{d-2}$ is the volume of the space with line element $d\Sigma_{(k) d-2}$; for $k=1$ this is simply the volume of a unit sphere in $(d-2)$-dimensions.

Next, to characterize the behaviour near the horizon, we proceed by expanding the metric function as
\be\label{eqn:nearHorizonFrobenius}
f(r) = 4 \pi T (r-r_+) + \sum_{i=2} a_n (r-r_+)^n \, ,
\ee
where $T = f'(r_+)/(4\pi)$ is the Hawking temperature.  We substitute this ansatz into the field equation~\eqref{EOM} and demand a solution order by order in $(r-r_+)$.  This produces, to second order, the following two constraints:
\ba \label{eqn:nearHorizon}
(d-2)C &=& (d-2) k r_+^{d-3} - \frac{2686}{13} \lambda k \delta_{d, 7}  - \frac{\lambda (d-3)}{2d-1}\times\nn\\
&&\times \Bigl[-\frac{{k} ( d-4 ) (
129-192d+{\frac {357}{4}}\,{d}^{2}-{\frac {57}{4}}\,{d}^{3}+{d}^{4}
 ) r_+^{d-7}}{3}\label{eqn:nearHorizon}
\nn\\
&&\quad+ (d^2 + 5d - 15) \bigl(64 k \pi^2 r_+^{d-5}  + \frac{256}{3} \pi^3 T r_+^{d-4}\bigr) T^2 \Bigr] \, ,
\\
0 &=& (d-2)(d-3)k r_+^{d-4} - 4 (d-2) r_+^{d-3} \pi  T - \frac{\lambda}{2d-1} \times\nn\\
&&\times\Bigl[
-\frac{{k}}{12}   \left( d-3 \right) \left( d-4 \right) \left( d-7 \right)
 \left( 516-768d+357{d}^{2}-57{d}^{3}+4{d}^{4} \right) r_+^{d-8}\nn\\
&&\quad -{\frac {128}{3}}{\pi }^{3} \left( d-4 \right)
 \left( d-3 \right)  \left( {d}^{2}+5\,d-15 \right) r_+^{d-5} T^3
\nn\\
&&\quad -64{\pi }^{2}  (d-3)(d-5) \left( {d}^{2}+5\,d-15 \right) kr_+^{d-6} T^2
\nn\\
&& \quad  +  \left( d-3 \right)  \left( d-4 \right)\left( d-6 \right)  \pi \left( 4{d}^{3}- 33 {d}^{2}+127 d- 166 \right) {k}^{2}
  r_+^{d-7} T  \Bigr]\,.\label{eqn:nearHorizon2}
\ea
These two equations allow one to determine $C$ (the mass) and $T$ in terms of the horizon radius, $r_+$.  Despite the fact that the above two equations arise from a series expansion near the horizon, we emphasize that the resulting expressions for the temperature and mass are exact.  Furthermore, the second equation, which determines the temperature, is a cubic equation; however only one of the roots will have a smooth $\lambda \to 0 $ limit.  It is this root which we consider in further analysis.

Although we have carried factors of $k$ to serve as accounting devices, we now set $k=1$ and determine if the first law of thermodynamics holds for the spherical, asymptotically flat black holes.  It is easiest to work with the expressions implicitly rather than attempting to solve the second equation for $T$.  To this end, $C$ can be traded in the first equation for the corresponding expression for the mass given in~\eqref{eqn:adm_mass}.  Varying both $M$ and $S$ with respect to $r_+$, we find that the first law of black hole thermodynamics
\be
\delta M = T \delta S \,
\ee
is satisfied.
Since we have computed the mass, entropy and temperature all independently, the fact that the first law holds provides a valuable check of our calculations.

One can also proceed to solve the field equation~\eqref{EOM} numerically: effectively joining the near horizon solution~\eqref{eqn:nearHorizonFrobenius} to the asymptotic solution~\eqref{eqn:asympExpansion}.  In practice, one must compute the near horizon solution to higher order than we have done in Eq.~\eqref{eqn:nearHorizon}, where only the first two terms were presented.  At order $(r-r_+)^3$, it is found that both $a_2$ and $a_3$ appear, with the latter appearing linearly.  Thus, $a_3$ can be solved for in terms of $a_2$.  This continues to arbitrary order, where at order $(r-r_+)^n$ one can obtain $a_n$ in terms of the previous coefficients.  Thus the near horizon solution is characterized by a single free parameter, $a_2$.  More explicitly we have
\be
a_2 = \frac{f''(r_+)}{2} = - \frac{(d-2)(d-3)}{r_+^2} \left[1 + \delta \right]
\ee
where $\delta$ characterizes the deviation from the ordinary Schwarzschild solution, characterized by  $\delta = 0$. To obtain the numerical solution, we first evaluated the near horizon series approximation close to the horizon.  In our calculations we have kept terms up to
order\footnote{This number of terms is on the excessive side, but the computation of the near horizon solution is easily automated, so keeping many terms in the expansion comes at no extra effort.} $(r-r_+)^{12}$. In practice, $\delta$ must be chosen very carefully to ensure consistency with the boundary condition $f({r \to \infty}) = 1$.    We employed the shooting method to determine $\delta$ and found that determining $\delta$ to approximately fifteen significant digits is sufficient to integrate the solution to approximately ten times the horizon radius.  At this point, the large $r$ series approximation becomes valid, and the solution can be continued to infinity in this way.
\begin{figure}[htp]
\centering
\includegraphics[width=0.4\textwidth]{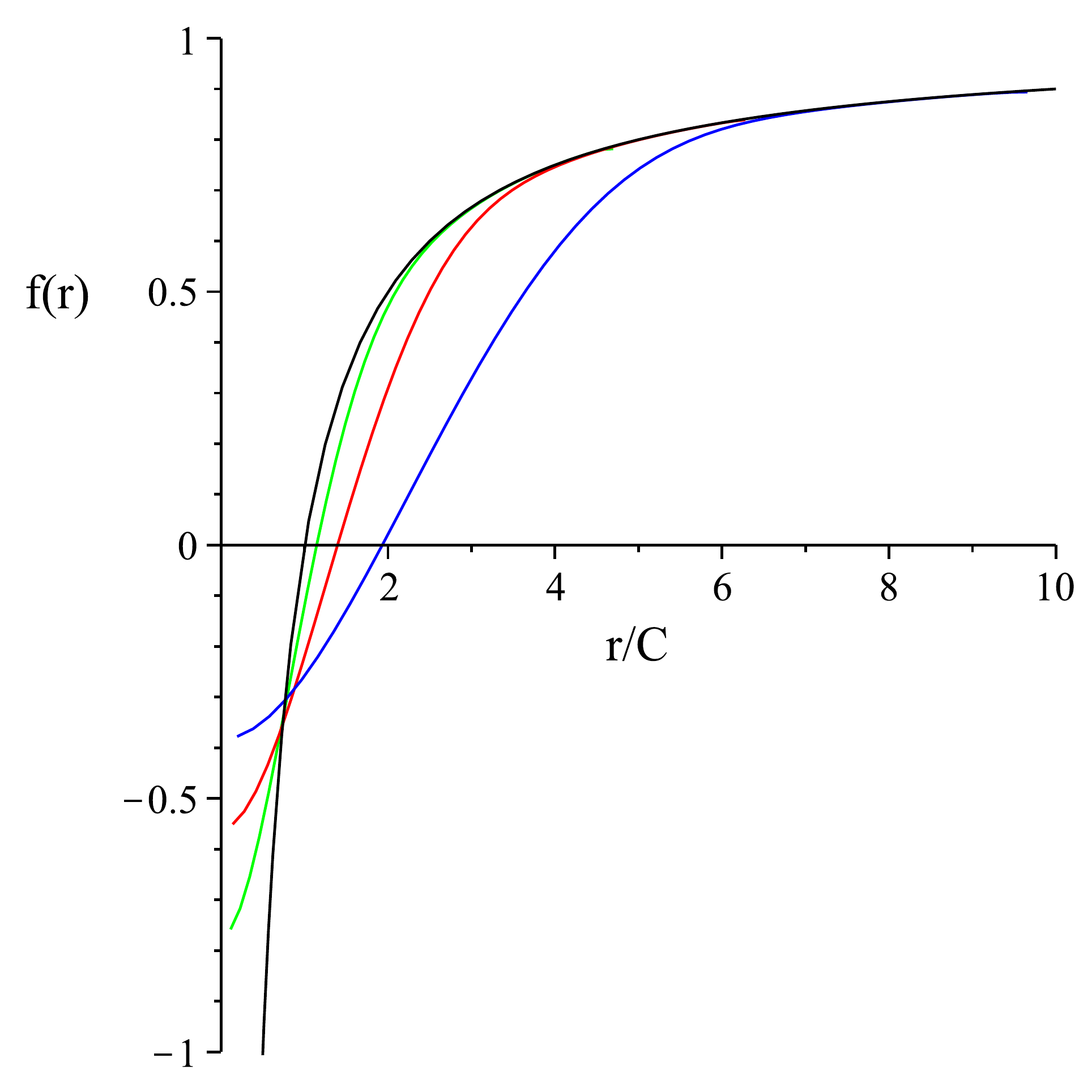}
\includegraphics[width=0.4\textwidth]{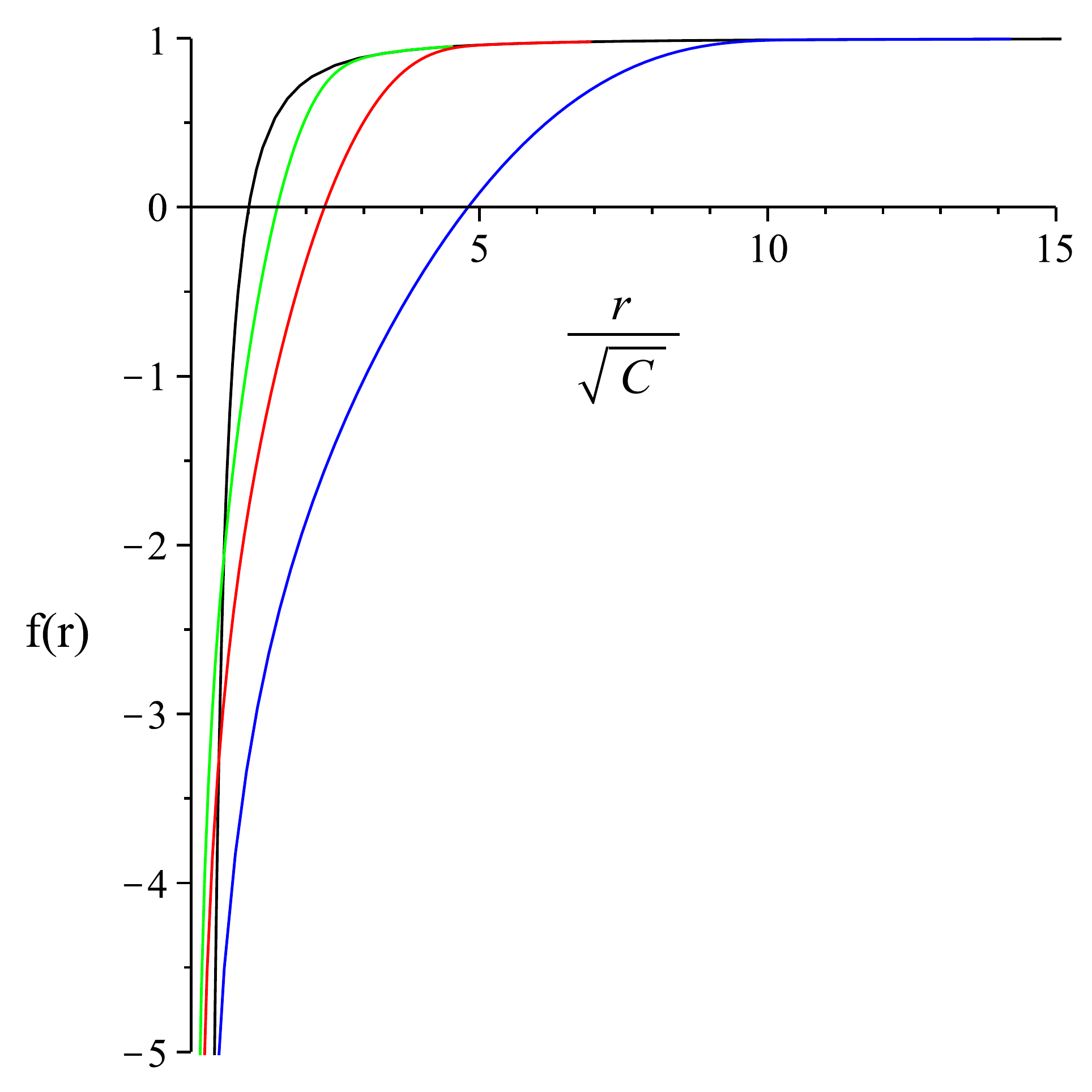}
\caption{{\bf Numerical solutions} (color online). \textit{Left}: $d=4$.  The black curve corresponds to the Schwarzschild solution.  The green, red and blue curves correspond to $\lambda/C_{d=4}^4 \approx 0.034, 0.237, 0.711$, respectively. \textit{Right}: $d=5$.  The black curve corresponds to the five dimensional Schwarzschild solution.  The green, red and blue curves correspond to $\lambda/C_{d=5}^2 \approx 0.040, 0.321, 6.822$, respectively. }
\label{fig:numerical_solns}
\end{figure}

The results of our numerical investigation are presented in Fig.~\ref{fig:numerical_solns}, where we display the solutions for four and five dimensions for various values of $\lambda$, made dimensionless via powers of $C$.  This demonstrates that it is possible to join the near horizon solution to the asymptotic expansion, confirming that   asymptotically flat black holes are indeed solutions of the theory. 
In general, stronger cubic coupling tends to push the horizon outward relative to the Schwarzschild case.  Note that, in the four dimensional case, the presence of the higher curvature terms softens the singularity: the Kretchmann scalar still diverges, but there is no metric singularity as $r \to 0$~\cite{Bueno:2016lrh}.
Of course, there remains much more to elucidate about the properties of these black holes, including the addition of matter as well as the study of AdS asymptotics and horizons of different topologies, but we shall leave this for future work.

\section{Conclusions}

We have studied gravitational actions containing terms up to cubic order in curvature in all dimensions and determined which of these theories admit natural extensions of the Schwarzschild solution, that is, admit a vacuum static spherically symmetric solution which can be characterized in terms of a single metric function.  We have demonstrated that to cubic order the most general theory  having this property takes the form \eqref{eqn:actionWithS}, where we identified a new term $S_d$, which has been overlooked in the literature to date.  This term has the remarkable property that it gives rise to a field equation that is a total derivative.  Therefore, in vacuum or in the presence of suitable matter (such as a Maxwell field), the equations of motion can be easily integrated once, resulting in the metric function satisfying a (nonlinear) second order differential equation.

We have also found the unexpected result that this class of theories has linearized equations of motion which are second order on a maximally symmetric background.  Further, the linearized equations on such a background coincide---up to an overall factor---with the linear Einstein equations on the same background.  As a result, the theory propagates only the massless, spin-2 graviton with the additional massive spin-2 and scalar modes absent.  It seems that such a feature is a necessary (but not sufficient) condition for a theory to permit a static, spherically symmetric solution described by a single metric function.  Though we have not found a rigorous proof of this result, we expect it to hold in higher order gravities as well.  Proving or disproving this statement is an interesting problem for future work.

Regarding the new theory arising from $\mathcal{S}_d$, there remain a number of questions to address.  First,
 while we have shown that the theory admits a Schwarzschild-like VSSS solutions, an interesting open question is whether the assumption on staticity can be relaxed; that is, does the Birkhoff theorem work for this theory, similar to Lovelock and quasitopological gravities~\cite{Oliva:2011xu, Oliva:2012zs, Ray:2015ava}?   Second, it should be possible to study black hole solutions which are asymptotic to AdS and study their extended thermodynamics~\cite{Kubiznak:2016qmn}.  Coupling the theory to Maxwell or scalar fields~\cite{Oliva:2011np} could lead to further examples of new, novel phase transitions such as the recently discovered superfluid-like transition~\cite{Hennigar:2016xwd}. It should also be possible to cast the theory in terms of horizon thermodynamics~\cite{Hansen:2016gud}, {at least for a certain class of matter actions, e.g. a Maxwell field}.    Furthermore, since the linearized equations coincide with the linearized Einstein equations, it should be possible to perform holographic studies of the theory, importing techniques which are now well developed~\cite{Myers:2010jv}.

\section*{Acknowledgments}

We are pleased to thank Rob Myers, Ruth Gregory, Jorma Louko and Eduardo Mart\'in-Mart\'inez for helpful comments and discussions.  This research was supported in part by Perimeter Institute for Theoretical Physics and by the Natural Sciences and Engineering Research Council of Canada. Research at Perimeter Institute is supported by the Government of
Canada through the Department of Innovation, Science and Economic
Development Canada and by the Province of Ontario through the
Ministry of Research, Innovation and Science.

\appendix

\section{Field equations}

In this appendix we present the full non-linear field equations 
for the theory \eqref{Sdtheory} minimally coupled to matter: 
\be
\mathcal{I} = \frac{1}{16 \pi G} \int d^d x \sqrt{-g} \Bigl(R -2\Lambda -  \lambda \mathcal{S}_d \Bigr) + \mathcal{I}_{\rm matter}\,.
\ee 
 The corresponding field equations can be most conveniently written in the following form:
\be
\mathcal{E}_{ab} = E_{acde} R_{b}{}^{cde} - \frac{1}{2} g_{ab} \mathcal{L} - 2 \nabla^c \nabla^d E_{acdb} = 8 \pi G T_{ab}\,,
\ee
where
\be
E^{abcd} = \frac{\partial \mathcal{L}}{\partial R_{abcd}}
\ee
is a complicated expression,  given by
\begin{align}
E_{abcd} &= \frac{1}{2} \left[  g_{ac} g_{bd} - g_{ad} g_{bc}  \right] - \frac{\lambda}{(d-2)(2d-1)} \bigg[
(4 d^2 +9d - 30) \left(  R_{ad} R_{bc} - R_{ac} R_{bd} \right)
\nn\\
&+ ( 2 d^2 - 35d + 66) \left( g_{ad} R_{b}{}^{e} R_{ce} -  g_{bd} R_{a}{}^{e} R_{ce}  + g_{bc} R_{a}{}^{e} R_{de}  -  g_{ac} R_{b}{}^{e} R_{de}\right)
\nn\\
&+ \frac{(4d^2 - 21d + 34)}{2} \left\{ \left(g_{ac} g_{bd} - g_{ad} g_{bc}  \right) R_{ef} R^{ef} +  \left(g_{bd} \
R_{ac} - g_{bc} R_{ad} - g_{ad} R_{bc} + g_{ac} R_{bd} \right) R\right\}
\nn\\
& - \frac{4d^2 - 13d + 30}{8} \left(g_{ac} g_{bd} - g_{ad} g_{bc} \right) R^2
\nn\\
&- \frac{4d^2 - 29d + 38}{8}  \left\{ 4 R R_{abcd} -  \left( g_{ac}g_{bd} - g_{ad}g_{bc} \right) R_{efhi} R^{efhi}  \right\}
\nn\\
&+  R_{d}{}^{e} R_{abce} -   R_{c}{}^{e} R_{abde} +  R_{b}{}^{e} R_{aecd} -   R_{a}{}^{e} R_{becd} - 21  R_{a}{}^{e}{}_{d}{}^{f} R_{becf}  + 21  R_{a}{}^{e}{}_{c}{}^{f} R_{bedf}
\nn\\
&+ (4d^2 + 9d - 30) \left(g_{bc} R^{ef} R_{aedf} - g_{bd} R^{ef} R_{aecf}  + g_{ad} R^{ef} R_{becf} - g_{ac} R^{ef} R_{bedf} \right)
\nn\\
&   + \frac{1}{2} \
 g_{bd} R_{a}{}^{efh} R_{cefh} \
-  \frac{1}{2}  g_{ad} R_{b}{}^{efh} R_{cefh} -  \frac{1}{2}  g_{bc} R_{a}{}^{efh} R_{defh} + \frac{1}{2}  g_{ac} \
R_{b}{}^{efh} R_{defh} \bigg] \, .
\end{align}
Due to the diffeomorphism invariance of the Lagrangian, the generalized Einstein tensor $\mathcal{E}_{ab}$ satisfies a Bianchi identity,
\be
\nabla^a \mathcal{E}_{ab} = 0 \, .
\ee

\bibliography{LBIB}
\bibliographystyle{JHEP}
\end{document}